\documentclass[superscriptaddress, aps, prd, reprint, nofootinbib]{revtex4-2}

\usepackage{graphicx}
\usepackage{tikz}
\usepackage{amsmath,amssymb,mathtools,mathrsfs,booktabs,multirow}
\usepackage{dcolumn}
\usepackage{bm}
\usepackage{xcolor}
\usepackage{soul}
\usepackage{braket}
\usepackage[separate-uncertainty=true,multi-part-units=single]{siunitx}
\usepackage{scalerel}
\usepackage{enumitem}
\usepackage{dsfont}
\usepackage{comment}

\definecolor{Mathematica1}{rgb}{0.368417, 0.506779, 0.709798}
\definecolor{Mathematica2}{rgb}{0.880722, 0.611041, 0.142051}
\definecolor{Mathematica3}{rgb}{0.560181, 0.691569, 0.194885}
\definecolor{darkred}{rgb}{0.545,0,0}
\definecolor{dullblue}{rgb}{0,0.298,0.49}
\definecolor{blue3}{RGB}{31, 119, 180}
\definecolor{andreapurple}{RGB}{130,0,180}
\usepackage[colorlinks=true
,urlcolor=dullblue
,anchorcolor=blue3
,citecolor=dullblue
,filecolor=blue3
,linkcolor=darkred
,menucolor=blue3
,pagecolor=blue3
,linktocpage=true
,pdfproducer=medialab
]{hyperref}

\makeatletter
\def\doauthor#1#2#3{%
  \global\let\@extramark\@empty
  \ignorespaces#1\unskip\@listcomma
  \begingroup
   #3%
  \@if@empty{#2}{\endgroup{}{}}{\endgroup{\comma@space}{}\frontmatter@footnote{#2}}%
  \@extramark
  \space \@listand
}%
\newcommand{\extranote}[1]{%
  \texorpdfstring{\gdef\@extramark{\comma@space\frontmatter@footnote{#1}}}{}%
}
\makeatother

\newcommand{\ped}[1]{_{\mathrm{#1}}}

\def\beq{\begin{equation}}
\def\eeq{\end{equation}}

\newcommand{\Msun}{M_\odot}

\newcommand{\Mbh}{M_\bullet}

\begin{document}

\title{Is S301 the Captured Companion of the Hypervelocity Star S5-HVS1?}

\author{Andrea Caputo}
\affiliation{Department of Theoretical Physics, CERN, 1217 Geneva, Switzerland}
\affiliation{Dipartimento di Fisica, ``Sapienza'' Università di Roma \& Sezione INFN Roma 1, 00185 Roma, Italy}
\affiliation{Department of Particle Physics and Astrophysics, Weizmann Institute of Science, Rehovot 7610001, Israel}

\author{Giovanni Maria Tomaselli}
\affiliation{School of Natural Sciences, Institute for Advanced Study, Princeton, NJ 08540, USA}

\begin{abstract}

Stellar binary disruptions through the Hills mechanism produce two fossils: a hypervelocity star (HVS), and a star tightly bound to the supermassive black hole. Among known galactic HVSs, only S5-HVS1 has unambiguous galactic-centre origin. Its measured mass and velocity determine a relation between the mass and semi-major axis of its captured companion. GRAVITY has now discovered S301, whose orbit and photometrically inferred mass satisfy this relation, making it the only compelling candidate for the captured companion of S5-HVS1. We build a forward model for the Hills origin and compare it to the null hypothesis. This confirms that S301's orbit aligns much more closely with that of S5-HVS1's companion than a typical S-star. However, the catalog-level Bayes factor remains of order unity and dependent on the probabilities of survival and detection. S301 is thus a compelling candidate, but establishing its association with S5-HVS1 will require improved mass measurements, chemical comparisons and GRAVITY-calibrated selection functions.

\end{abstract}

\maketitle

\emph{Motivation.} When a stellar binary passes sufficiently close to a massive black hole, it can be tidally separated in a process known as ``Hills mechanism''~\cite{Hills:1988cvl, Yu:2003hj}. One of the two stars is left bound to the black hole on a tight, highly eccentric orbit, while its companion is ejected with a speed that may exceed the Galactic escape velocity, thus becoming a hypervelocity star (HVS). At the center of the Mikly Way, the $\sim\!\num{4e6}M_\odot$ black hole Sgr A* creates the environment needed for this process to occur. Binary disruptions near Sgr A* could account for both HVSs observed in the galactic galo, and part of the stars (known as S-stars) populating the central $\sim\!\SI{40}{mpc}$. These two populations are however usually studied separately, and no HVS has so far been associated to its bound companion. Identifying one such connection would allow us to reconstruct the close encounter that produced both fossils.

Most confirmed or candidate HVSs have uncertain birthplaces. Among them, S5-HVS1 is exceptional, because its trajectory traces securely back to the galactic center with a flight time $t_f\simeq\SI{4.8}{Myr}$, suggesting it originated through the Hills mechanism~\cite{Koposov2020}. This hypothesis can be turned into a quantitative prediction for the properties of its captured companion---see~\cite{Lu2021} for a detailed analysis. Let $m_e$ and $m_b$ denote the mass of the ejected and bound star, and $\epsilon_e$ and $\epsilon_b$ their current energies. Neglecting the binary's initial kinetic and potential energies, conservation of energy gives $m_e\epsilon_e+m_b\epsilon_b=0$, hence
\beq
a_b=\frac{G\Mbh}{m_ev_\infty^2}\,m_b\approx\SI{500}{AU}\,\bigg(\frac{m_b}{M_\odot}\bigg)\,,
\label{eq:line}
\eeq
where $\Mbh=\num{4.30e6}M_\odot$ is Sgr A*'s mass~\cite{GRAVITY2022}, $m_e=\num{2.35+-0.06}M_\odot$ is S5-HVS1's mass (estimated through a combination of spectroscopic and photometric measurements), $v_\infty=\SI{1799}{km/s}$ is its velocity~\cite{Koposov2020}, and $a_b$ and $m_b$ are the (initial) semi-major axis and mass of its captured companion. These last two parameters must therefore lie on the line defined by Eq.~\eqref{eq:line}. This relation has an intrinsic uncertainty of approximately $3\%$, dominated by the error on $m_e$, plus similar systematics due to the unknown initial binary's energy. 

The usefulness of this relation is that these parameters of the two stars do not change substantially over time. Both stars likely remain on the main sequence during the flight time, so their masses change negligibly. Furthermore, the relaxation timescale $t_E$ of the energy (and thus semi-major axis) of the captured companion due to two-body relaxation is long. Specifically, for the parameters of interest, cusp models consistent with the GRAVITY bound on the mass enclosed within S2's orbit give $t_E\gtrsim\SI{1}{Gyr}$~\cite{Hopman:2006qr,binney1987,GRAVITY2022}. Conservatively adopting this value, the induced semi-major axis diffusion is of the order $\Delta a_b/a_b\sim\sqrt{t_f/t_E}\simeq 7\%$.

While the mass and semi-major axis are robust fossils of the Hills encounter, coherent torques lead to a much faster angular momentum diffusion. Its magnitude (and thus the orbital eccentricity) evolves under the action of scalar resonant relaxation, while its direction (and thus the orbital plane) changes due to the even faster vector resonant relaxation. The time evolution of these quantities must therefore be explicitly modeled.

\emph{S301: a new candidate.} The GRAVITY collaboration has recently discovered S301, a faint star on a $\num{8.7}$-year orbit with semi-major axis $a=\SI{687+-6}{AU}$ and eccentricity $e=0.9832\pm0.0010$~\cite{AbdElDayem2026}.\footnote{The orbit can also be fitted with mirrored orientation and eccentricity $e=0.9824\pm0.0011$~\cite{AbdElDayem2026}.} At this semi-major axis, Eq.~\eqref{eq:line} predicts a mass
\beq
m_b=\bigg(\frac{a_b}{\SI{500}{AU}}\bigg)\Msun\simeq1.37\,\Msun\,.
\eeq
This value lies within the $1.1$--$1.5M_\odot$ range inferred from S301's $K$-band magnitude $m_K=19.3$ using main-sequence stellar
models and MIST/PARSEC isochrones~\cite{AbdElDayem2026, Pecaut:2013kka}. Within the present photometric uncertainty, S301 thus sits on the mass-energy relation required to be the former companion of S5-HVS1.

Its orbit provides a second, complementary cue. With a pericenter of just $r_p=a(1-e)\simeq\SI{11.5}{AU}$, S301 has the closest known approach to Sgr A* of any other S-star. Such an orbit is qualitatively expected after a Hills disruption. Indeed, tidal disruption requires the original binary to approach Sgr A* within its tidal radius $r_p\sim r_t=a\ped{bin}(\Mbh/m\ped{bin})^{1/3}$, where $a\ped{bin}$ and $m\ped{bin}$ are the semi-major axis and mass of the original binary. At breakup, the tidal energy spread between the two stars is $\Delta\epsilon\sim G\Mbh a\ped{bin}/r_t^2$, about half of which equals the eventual orbital energy of the captured star. We thus have
\beq
1-e=\frac{r_p}{a}\sim\frac{r_t}{G\Mbh/\Delta\epsilon}\sim\bigg(\frac{m\ped{bin}}{\Mbh}\bigg)^{1/3}\sim10^{-2}\,.
\label{eqn:initial-e}
\eeq
This prediction naturally includes an order-unity ambiguity encoding the actual center-of-mass approach distance, binary phase, binary internal parameters, etc.

As anticipated, however, eccentricity diffusion is generally not negligible over the flight time of S5-HVS1. For a S301-like orbit, Ref.~\cite{AbdElDayem2026} finds a diffusion time of
the dimensionless angular momentum $j\equiv\sqrt{1-e^2}$ of
$t_j\simeq(2$--$3)\times10^7\,\si{yr}$. This estimate is dominated by
non-resonant two-body relaxation, whose diffusion coefficient
$D_{jj}=j_{301}^2/t_j$ is independent of $j$ (resonant relaxation being quenched by the rapid Schwarzschild precession). The angular momentum spread accumulated over the flight time is thus
\beq
\Delta j\simeq\sqrt{D_{jj}t_f}=j_{301}\sqrt{\frac{t_f}{t_j}}\simeq0.07\text{--}0.09\,,
\eeq
where $j_{301}\simeq0.18$ is the dimensionless angular momentum of S301's measured orbit,
at which $t_j$ is evaluated. Applied to the initial eccentricity
\eqref{eqn:initial-e}, this gives $e\gtrsim0.95$ today.

\emph{Other stars.} One may ask whether any other S-star also satisfies the freshness condition and the energy relation \eqref{eq:line}. The former is sufficient to cut off most candidates. For a power-law stellar density profile $\rho\propto r^{-\gamma}$, the diffusion time scales as $t_j\propto a^{\gamma-3/2}$. Extrapolations from the observed diffuse light within $r\lesssim\SI{0.5}{pc}$ from galactic center give slopes as low as $\gamma\simeq1.1$~\cite{2018A&A...609A..27S}, while a fully relaxed Bahcall-Wolf cusp has $\gamma=7/4$~\cite{1976ApJ...209..214B}. Across this range, $t_j$ varies by at most a factor of order unity for different S-Stars, so the cut quoted above can be robustly applied to the whole catalog. Among the 39 known bound orbits~\cite{2017ApJ...837...30G,GRAVITY:2024tth}, only three stars have $e>0.93$: S14, S29, and S175, with eccentricities $0.9761$, $0.9688$, and $0.9867$, and semi-major axes $\SI{2380+-30}{AU}$, $\SI{3230+-15}{AU}$ and $\SI{3450+-330}{AU}$ respectively.\footnote{Reference~\cite{2020ApJ...899...50P} also reported two highly eccentric stars, S62 and S4714, but subsequent measurements by the GRAVITY collaboration failed to reproduce their results \cite{2021A&A...645A.127G,2022A&A...657A..82G}.}

We can then use the energy relation to discriminate between these three and S301, although this requires dealing with the large uncertainties on photometric mass estimates, as none of these stars has a spectroscopic determination of the mass. Their $K$-band magnitudes $m_K=15.7$, $16.7$, and $17.5$ imply masses of roughly $7.5$--$8.7M_\odot$, $4.5$--$5.1M_\odot$, and $2.9$--$3.4M_\odot$, respectively, where we applied the updated $K$-band main-sequence relation from Ref.~\cite{Pecaut:2013kka,Mamajek:2022dwarf}, adopting a foreground extinction of $2.4$--$2.7$ magnitudes (for S301, this procedure gives $1.3$--$1.5\,M_\odot$, consistent with the MIST/PARSEC isochrone fit of Ref.~\cite{AbdElDayem2026}). On the other hand, Eq.~\eqref{eq:line} would require $4.8M_\odot$, $6.5M_\odot$, and $7M_\odot$, respectively, none of which lies within the corresponding photometric interval. Systematic errors are possible, but they are unlikely to reconcile S175, which is $1.4$--$1.7$ magnitudes fainter than required, whereas S29 is $0.5$--$0.8$ magnitudes fainter and is therefore disfavored but less securely excluded. On the other hand, across bright S-stars with spectroscopic--evolutionary mass estimates, the $K$-band relation tends to overestimate the mass by about $10\%$, with offsets reaching $30\%$ (e.g., $17.3M_\odot$ against $13.6^{+2.2}_{-1.8}\,\Msun$ for S2~\cite{Habibi_2017}). Although this correction is not by itself sufficient to bring S14 onto the line, it pushes the star towards Eq.~\eqref{eq:line}, so S14 can be disfavored but not definitely excluded until a spectroscopic measurement becomes available.

Even with this caveat, we conclude that, among the known stars, only S301 cleanly satisfies \emph{both} the energetic requirement and the fresh-capture eccentricity.

\textit{Statistical tests.} The uniqueness of S301 does not give, by itself, any probabilistic information regarding its association with S5-HVS1. We can give such quantification through the Bayes factor
\beq
\mathcal B\ped{301}\equiv\frac{f\ped{pair}^{(301)}(a,e)}{f_0^{(301)}(a,e)}\,,
\label{eqn:bayes}
\eeq
where $f\ped{pair}$ and $f_0$ are the orbital parameter likelihoods under the hypothesis of pairing with S5-HVS1, and under the null hypothesis, evaluated at S301's orbital parameters. The calculation could be refined by including the mass distribution in the likelihood. We prefer however to keep it out of our estimates given the large photometric mass uncertainties discussed above.

We build the likelihoods as follows (see also~\cite{Lu2021} for a similar approach). For the pairing hypothesis, we draw a primary mass $m_1$ from Kroupa's mass function~\cite{Kroupa:2000iv}, and a secondary mass $m_2=qm_1$ with the mass ratio distributed as $p(q)\propto q^{-0.3}$ on $[0.1,1]$ with a $10\%$ twin excess uniform in $[0.9,1]$~\cite{Moe:2017icj}. We assign equal prior probability for either star to be ejected, weighting the results by a gaussian likelihood corresponding to S5-HVS1's measured mass, $\num{2.35+-0.06}M_\odot$. The companion mass $m_b$ is that of the other member. Finally, we restrict to $m_b\ge1.1M_\odot$, as lower masses would be invisible by GRAVITY (having a $K$-band magnitude $m_K\gtrsim20$), and to $m_b<7M_\odot$, because the lifetime of more massive stars would be shorter than S5-HVS1's age of $\sim\!\SI{50}{Myr}$. Having sampled $m_b$, we map it to the semi-major axis $a$ using $\eqref{eq:line}$. Finally, we assign an initial eccentricity from the capture relation \eqref{eqn:initial-e}, promoting its order-unity ambiguity to a random variable: we set $1-e=\kappa\,(m\ped{bin}/\Mbh)^{1/3}$ with $m\ped{bin}=m_e+m_b$, and draw $\kappa$ from a lognormal distribution with median $1.5$ and $\sigma_{\ln\kappa}=0.45$. This choice is calibrated to reproduce the eccentricities of captured stars in the Galactic-center binary-disruption simulations of Ref.~\cite{Generozov2025}. Each star is then evolved for \SI{4.8}{Myr} using random walk angular momentum diffusion with $t_j=(\SI{3e7}{yr})(a/a_{301})^{\gamma-3/2}$, and an absorbing boundary at $\SI{3}{AU}$, below which cumulative tidal heating disrupts the star within the flight time~\cite{Lu2021}.

For the null hypothesis likelihood $f_0(a,e)$, we assume a thermal distribution $p(e)=2e$. For its semi-major axis dependence we need the slope $\gamma$ of the density profile $\rho\propto r^{-\gamma}$. We calculate the results for the same range of $\gamma$ used earlier, spanning extrapolations from diffuse light, $\gamma=1.1$~\cite{2018A&A...609A..27S}, and theoretical predictions, $\gamma=1.75$~\cite{1976ApJ...209..214B}. Finally, we normalize $f_0$ over the same interval of semi-major axes where $f\ped{pair}$ has support, namely $a\in[550,3500]\,\si{AU}$. The Bayes factor $\mathcal B\ped{S301}$ can then be computed as in \eqref{eqn:bayes}.

The number obtained this way, however, suffers from the look-elsewhere effect: a high Bayes factor is expected for the single star whose orbital parameters are a best match for the pairing hypothesis. We can correct for this by computing instead the Bayes factor for the hypothesis that the \emph{catalog} of the $N=18$ stars with $a\in[550,3500]\,\si{AU}$ contains the captured companion, against the null hypothesis that all stars are unrelated---for a similar strategy, see~\cite{1992MNRAS.259..413S, Ashton:2017ykh}. If $p\ped{det}$ is the probability that S5-HVS1's companion survives for a time $t_f$ and is detected in the catalog, we have
\beq
\begin{split}
\mathcal B\ped{cat}&\equiv\frac{(1-p\ped{det})\prod_{i=1}^Nf_0^{(i)}+\frac{p\ped{det}}{N}\sum_{i=1}^Nf\ped{pair}^{(i)}\prod_{j\neq i}f_0^{(j)}}{\prod_{j=1}^Nf_0^{(j)}}\\
&=(1-p\ped{det})+\frac{p\ped{det}}{N}\sum_{i=1}^N\mathcal B_i\,,
\end{split}
\label{eq:BayesFinal}
\eeq
where $f^{(i)}$ denotes the likelihoods evaluated at the parameters of the $i$-th star, and $\mathcal B_i$ is its individual Bayes factor, defined as in \eqref{eqn:bayes}. This calculation treats the catalog entries as exchangeable: conditional on the companion being detected, each of the $N$ stars has the same prior probability $1/N$ of being it. Their measured properties then distinguish them through the individual Bayes factors $\mathcal B_i$.

We release on GitHub \cite{S301_S5-HVS1_Bayes:github} scripts that can be used to reproduce our model and results. For the two choices of density slope, we obtain
\begin{center}
\setlength{\tabcolsep}{12pt}
\begin{tabular}{rccc}
& $\gamma=1.1$ & $\gamma=1.75$\\
\midrule
$\mathcal B_{301}=$ & $90$ & $42$\\
$\mathcal B_{14}=$ & $2.4$ & $3.2$\\
$\mathcal B_{29}=$ & $0.58$ & $0.90$\\
$\mathcal B_{175}=$ & $0.59$ & $1.1$\\
$\mathcal B\ped{cat}=$ & $1+4.2\,p\ped{det}$ & $1+1.6\,p\ped{det}$\\
\end{tabular}
\end{center}
where stars other than the ones mentioned here contribute negligibly. The large Bayes factor $\mathcal B_{301}$ confirms that S301's orbit aligns much more closely with that of S5-HVS1's companion than a typical S-star. On the other hand, the catalog-level $\mathcal B\ped{cat}$ remains modest (of order unity) while still favoring the pair hypothesis for any nonzero $p\ped{det}$.\footnote{To give a rough idea, on conventional descriptive scales~\cite{kass1995bayes}, a Bayes factor of three would lie near the boundary between evidence ``not worth more than a bare mention'' and positive or substantial evidence} The bare survival chance was estimated as $5$--$50\%$ by Ref.~\cite{Lu2021}, while a quantitative estimate of $p\ped{det}$ requires combining a physical forward model with the selection function of GRAVITY's observations. The latter ingredient can only be provided by the GRAVITY Collaboration itself, e.g., by injection and recovery of artificial stars. This would also allow to refine the equal $1/N$ weights used in Eq.~\eqref{eq:BayesFinal}.

\emph{Outlook.} Future spectroscopy, for example with the Extremely Large Telescope~\cite{ELT2018}, can provide the most decisive tests. A precise mass measurement could reduce the uncertainty enough to either pin S301 on the energy line~\eqref{eq:line} or away from it. Furthermore, trustworthy masses can be factored in the likelihood and make the statistical case stronger. On top of that, co-natal former binary companions should share multi-element abundance patterns~\cite{2019ApJ...871...42A, Hawkins_2020}, which are much harder to obtain accidentally than agreement on metallicity alone. Combining a catalog-level likelihood with a matching age and chemical fingerprint could either reject the association, or make S301--S5-HVS1 the first identified ejected-captured Hills pair.

\textit{Acknowledgments} A.C. is supported by an ERC STG grant (“AstroDarkLS”, grant No. 101117510). G.M.T. gratefully acknowledges support from the Rubicon Fellowship, awarded by the Netherlands Organisation for Scientific Research (NWO), Grant ID
\href{https://doi.org/10.61686/WYKDB06497}{https://doi.org/10.61686/WYKDB06497}. AC also acknowledges the Weizmann Institute of Science for hospitality during this project and the support from the Benoziyo Endowment Fund for the Advancement of Science.

\bibliography{main}

\end{document}